\documentclass[nofootinbib,superscriptaddress]{revtex4}

\usepackage{amsmath}
\usepackage{graphicx}

\def\GeV{{\rm\ GeV}}
\def\ve{\varepsilon}
\def\CG{{\cal G}}
\def\CM{{\cal M}}
\def\be{\begin{equation}}
\def\ee{\end{equation}}
\def\bea{\begin{eqnarray}}
\def\eea{\end{eqnarray}}
\def\Re{\mathop{\rm Re}\nolimits}
\def\Im{\mathop{\rm Im}\nolimits}

\begin{document}

\title{On $\Delta$ resonance contribution to two-photon exchange amplitude}

\author{Dmitry~Borisyuk}
\affiliation{Bogolyubov Institute for Theoretical Physics,
14-B Metrologicheskaya street, Kiev 03680, Ukraine}

\author{Alexander~Kobushkin}
\affiliation{Bogolyubov Institute for Theoretical Physics,
14-B Metrologicheskaya street, Kiev 03680, Ukraine}
\affiliation{National Technical University of Ukraine "KPI",
37 Prospect Peremogy, Kiev 03056, Ukraine}

\begin{abstract}
 We consider two-photon exchange (TPE)
 in the elastic electron-proton scattering
 and study the contribution arising from the production of
 $\Delta(1232)$ resonance in the intermediate state.
 We calculate all three TPE amplitudes (generalized form factors),
 and find that the $\Delta$ contribution mainly influences
 generalized electric form factor
 (contrary to the elastic contribution, which affects magnetic form factor),
 and the effect grows with $Q^2$.
 If the corresponding correction is applied to the recent
 polarization transfer measurements of proton form factors,
 their results will change markedly.
 Thus we suggest that TPE corrections due to inelastic intermediate states
 are important to polarization experiments at high $Q^2$,
 and should not be neglected.
\end{abstract}

\maketitle

\section{Introduction} 
\begin{figure}
 \includegraphics[width=0.2\textwidth]{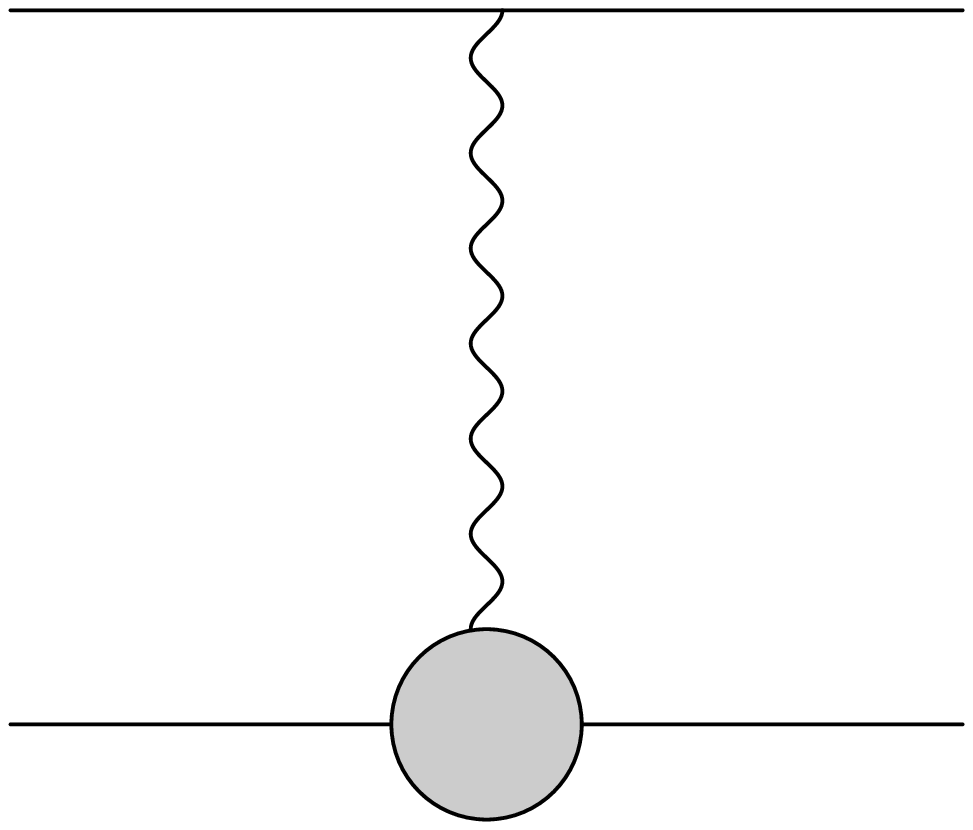}\qquad\qquad
 \includegraphics[width=0.2\textwidth]{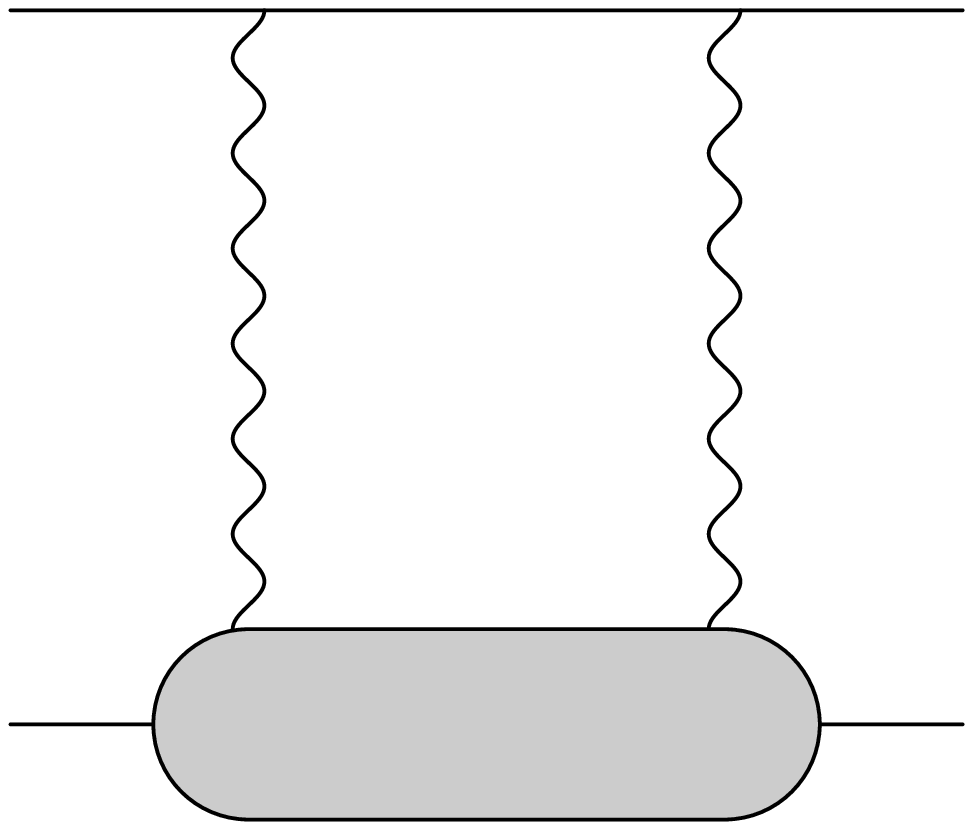}
 \caption{One- and two-photon exchange diagrams} \label{TPEdiagr}
\end{figure}

Due to smallness of the fine structure constant $\alpha\approx\frac{1}{137}$,
the elastic electron-proton scattering amplitude is dominated
by the first-order term, corresponding to the exchange
of a single photon, Fig.~\ref{TPEdiagr}, left.
The one-photon exchange (OPE) amplitude has a specific structure,
which allows, e.g., for the Rosenbluth separation of form factors.
In the next (second) order the only non-trivial diagram
is two-photon exchange (TPE), Fig.~\ref{TPEdiagr}, right.
%
Despite its smallness, in some cases TPE correction is very important,
because it changes qualitatively the structure of the scattering amplitude.
Thus, TPE influence naturally explains the discrepancy between Rosenbluth
and polarization methods in proton form factor (FF) measurements.
For a further review and up-to-date bibliography see e.g. Ref.~\cite{Review}.

The full TPE amplitude may be split into separate contributions
according to hadronic intermediate state (IS),
which depicted as the blob in TPE diagram.
We will have the elastic contribution (pure proton IS),
and inelastic ones, which come from multi-particle states
such as $p\pi$, $p\pi\pi$, $p\eta$ and so on.
We may also distinguish the contributions of resonances,
such as Delta, Roper etc.
(This is so-called "hadronic approach".
There are also QCD-style calculations \cite{parton,ourQCD},
which assume IS to be a set of quarks.
In this paper we resort to the former approach).


At first, the TPE amplitude was approximated by the elastic contribution,
which is the most well-studied one.
Then, it was proposed to study contributions of different
hadronic resonances as ISs.
Kondratyuk et al. performed calculations with Delta \cite{BlundenDelta}
and several other light resonances \cite{BlundenRes}.
They studied the TPE correction to the cross-section and concluded that
Delta contribution is the largest among all resonances,
though still much smaller than the elastic one.
Similar results were obtained for target normal spin asymmetry,
the observable that is related to imaginary part
of the TPE amplitude \cite{ourTNSA}.

In this paper we present new results for Delta resonance contribution
to the TPE amplitude in the elastic $ep$ scattering.
To calculate it we employ the dispersion method, which was
developed in Ref.~\cite{ourDisp} for $ep$ scattering
and applied to $e\pi$ scattering in Ref.~\cite{ourPi}.
The method is described in detail in those papers,
here we just recall that it
\begin{itemize}
 \item ensures correct behaviour of the TPE amplitude at $\ve\to 1$
   (the amplitude goes to zero),
 \item eliminates a need for off-shell FFs.
\end{itemize}
In comparison to previous works of Kondratyuk et al. \cite{BlundenDelta,BlundenRes},
we consider not just cross-section correction, but all three generalized FFs
(TPE amplitudes) and discuss corrections to polarization transfer (PT)
experiments, which was never done previously.

\section{Theoretical background}
To describe the elastic scattering amplitude in the presence of TPE,
we will use the amplitudes $\CG_E$, $\CG_M$ and $\CG_3$,
defined in Ref.~\cite{ourDisp}:
\be
 \begin{array}{ccl}
  \CG_E & = & \tilde F_1 - \tau \tilde F_2 + \nu \tilde F_3/4M^2\\
  \CG_M & = & \tilde F_1 +\tilde F_2 + \ve \nu \tilde F_3/4M^2\\
  \CG_3 & = & \nu \tilde F_3/4M^2
 \end{array}
\ee
where $\tilde F_i$ are related to the scattering amplitude as
\be \label{TPEampl}
  \CM_{fi} = -\frac{4\pi\alpha}{q^2} \bar u'\gamma_\mu u \cdot
 \bar U' \left(\gamma^\mu \tilde F_1
 - \frac{1}{4M} [\gamma^\mu,\hat q] \tilde F_2
 + \frac{P^\mu}{M^2} \hat K \tilde F_3
 \right) U
\ee
and all notation is identical to that of Ref.~\cite{ourDisp}.
The TPE contribution will be indicated by the prefix $\delta$, viz.
\be
 \CG_E = G_E + \delta\CG_E
       = G_E + \delta\CG_E^{(el)} + \delta\CG_E^{(\Delta)} + ...
\ee
where $G_E$ is usual proton FF,
$\delta\CG_E^{(el)}$ is elastic contribution,
$\delta\CG_E^{(\Delta)}$ is Delta resonance contribution,
and the contributions of other ISs (neglected hereafter),
are indicated by the ellipsis.
As the amplitude $\CG_3$ is absent in OPE approximation,
it coincides with the corresponding TPE contribution, $\CG_3 = \delta\CG_3$.

Recall that the observables are expressed via these amplitudes as follows.
The reduced cross-section correction is
\be
 \delta\sigma_R = 2 \Re \left( \ve G_E \delta\CG_E + \tau G_M \delta\CG_M \right)
\ee
and the main contribution to it comes from $\delta\CG_M$ \cite{ourPheno},
the correction to the FF ratio%
\footnote{%
 What is measured in polarization experiments is proton L/T polarization ratio,
 which, {\it in OPE approximation}, is proportional
 to $G_E/G_M$ ratio. It will be convenient to divide polarization ratio
 by the appropriate kinematical factor and define
 {\it experimentally measured} FF ratio $R$, which equals to $G_E/G_M$
 in OPE approximation but really differs because of TPE corrections.
}
 $R=G_E/G_M$ is
\be \label{dR}
 \delta R = R \, \Re \left( \frac{\delta\CG_E}{G_E} - \frac{\delta\CG_M}{G_M}
    - \frac{\ve(1-\ve)}{1+\ve} \, \frac{\delta\CG_3}{G_M} \right)
\ee
and the last term in the brackets has little effect because of small factor
$\tfrac{\ve(1-\ve)}{1+\ve}$.

It is interesting to note that the same combination of the amplitudes
determines target normal spin asymmetry \cite{ourTNSA}:
\be \label{TNSA}
 A_n = - \sqrt{2\ve(1+\ve)} \frac{2QMR}{Q^2 + 4M^2 R^2 \ve}
 \Im \left( \frac{\delta\CG_E}{G_E} - \frac{\delta\CG_M}{G_M}
     - \frac{\ve(1-\ve)}{1+\ve} \, \frac{\delta\CG_3}{G_M} \right).
\ee

The $\Delta N\gamma^*$ vertex, in general, contains three FFs
(magnetic, electric, and Coulomb ones).
However, it is well-known, that electric and Coulomb FFs
are small and experimental data
are described rather well by single magnetic FF.
Thus, to simplify the calculation, we assume purely magnetic
$\Delta \to N\gamma^*$ transition,
which may be described by the following amplitude (see e.g. \cite{DeltaFF}):
\be
 \CM_{\Delta N\gamma} = \sqrt{4\pi\alpha} \, i
    \epsilon^{\mu\alpha\beta\gamma} p_\beta q_\gamma \, \bar U V_\alpha \,
    \frac{F_\Delta(q^2)}{2M^2}
\ee
where $p$ and $q$ are Delta and photon momenta, respectively
(the nucleon momentum thus will be $p-q$), $U$ is nucleon spinor,
and $V_\alpha$ is Rarita-Schwinger wavefunction of Delta resonance,
normalized according to $\bar V_\alpha V_\alpha = - 2 M_\Delta$.
The transition FF was expressed as
\be \label{DeltaFF}
 F_\Delta(q^2) = \sum_{i=1}^5 \frac{c_i q^2}{q^2-m_i^2}
\ee
with $m_i$ and $c_i$ given in Table~\ref{Tab:DeltaFF}.
\begin{table}
\begin{tabular}{|c@{\quad}|*{5}{l@{\quad}|}}
\hline
$m_i$ & 0. & 2.170270 & 0.660810 & 0.715202 & 0.768494\\
\hline
$c_i$ & -3.377428 & 0.072839 & -20.794000 & 69.497989 & -45.399399\\
\hline
\end{tabular} \caption{Parameters of $N\to\Delta$ transition form factor, Eq.(\ref{DeltaFF}).}\label{Tab:DeltaFF}
\end{table}
These values were obtained by fitting experimental data
from Ref.~\cite{Stoler}. 
We will neglect the width of the resonance,
as it was done in Ref.~\cite{BlundenDelta}.

For the elastic proton FFs we use parameterization from Ref.~\cite{Arrington}.

\section{Results and discussion}
\begin{figure}
 \includegraphics[width=0.48\textwidth]{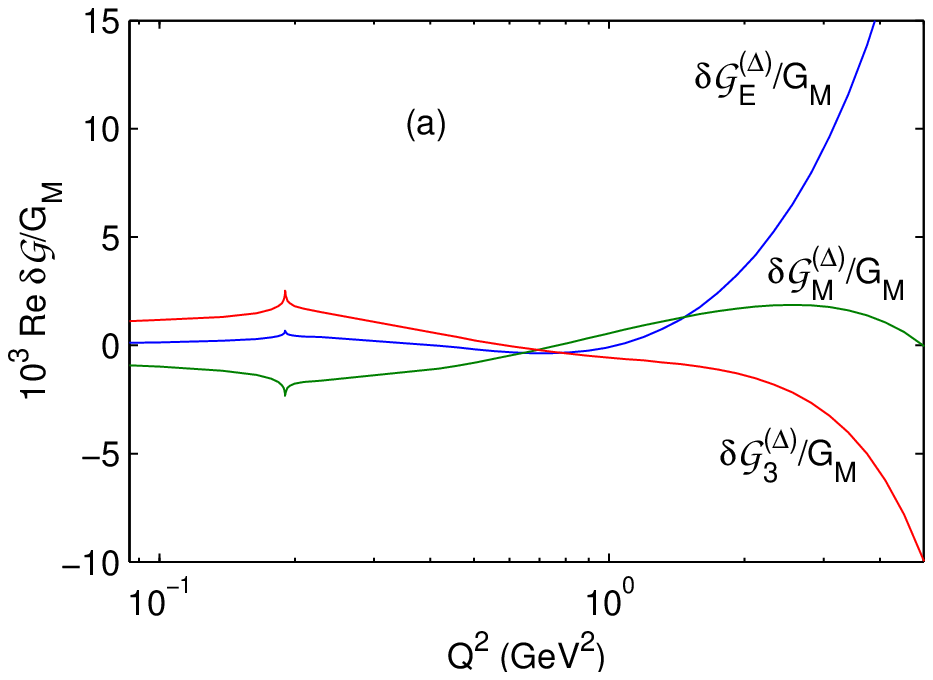}\quad
 \includegraphics[width=0.48\textwidth]{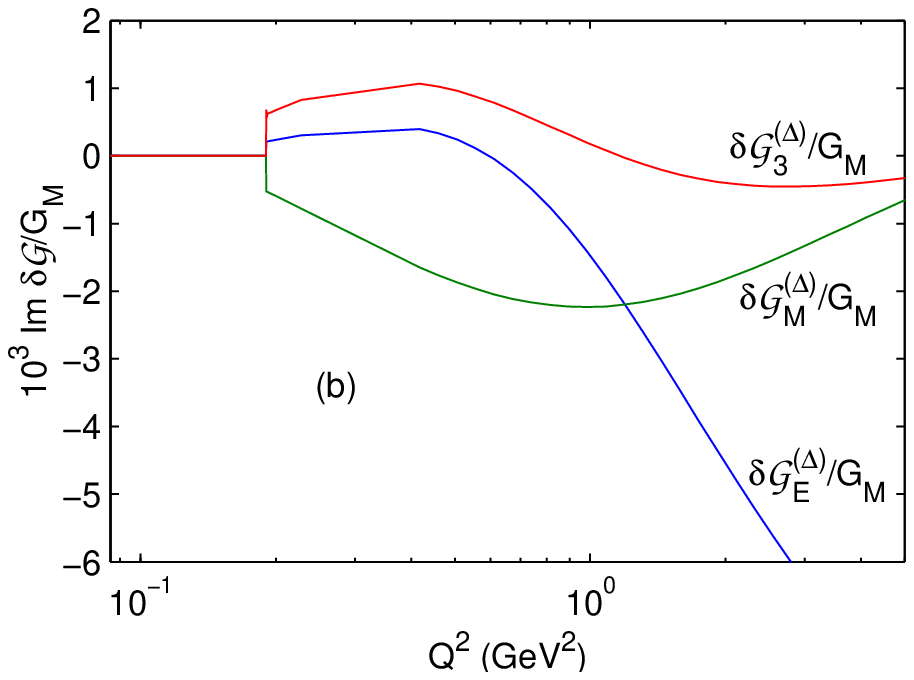}
 \caption{(Color online) Delta resonance contribution to TPE amplitudes, $\ve=0.25$, real part (a) and imaginary part (b).}\label{deltaG} 
\end{figure}
Figure~\ref{deltaG} displays $Q^2$ dependence of Delta resonance contribution
to the TPE amplitudes at fixed $\ve$ ($\ve=0.25$).
As usual, we consider "normalized" TPE amplitudes
(divided by the proton magnetic FF).
The scale is made logarithmic in $Q^2$ for better display of low-$Q^2$ region.

Looking at the real parts of the amplitudes, we see sharp peaks
coinciding with the resonance position.
It was noted in Ref.~\cite{ourPi}, that such peaks are artefacts, appearing
due to assumed zero resonance width.
With a finite width, the curve must become "smeared" and the peaks
should disappear.
But this means that the TPE amplitude in the close vicinity of resonance 
is not adequately described by the present "zero-width" calculation.
Further we will mainly concentrate on high-$Q^2$ region ($Q^2 > 1 \GeV^2$),
where we do not hit the resonance and the problem will not emerge.


We see that at high $Q^2$ Delta contributions grow (in absolute value)
with $Q^2$. Though the contribution $\delta\CG^{(\Delta)}_M/G_M$
changes sign at $Q^2\approx 5\GeV^2$, it still grows beyond
this point (not shown in the figure).
The elastic contribution has similar property (Fig.~\ref{vs.elastic}).
The difference is that the largest contribution goes to the amplitude
$\CG_E$ (much more larger than to $\CG_M$).
This fact has not much effect on the cross-section,
but implies relatively large corrections to the polarization ratio (see below).

The imaginary parts (Fig.~\ref{deltaG}b),
naturally, have a step-like behaviour,
i.e. they vanish below the threshold and are non-zero above it.
Having obtained the imaginary part of the amplitudes,
we can perform some cross-checks of our results.
First, we can check the sign of the TPE amplitudes
with the help of optical theorem.
It reads
\be
 \Im \CM_{ii} = 2|p|\sqrt{s} \, \sigma
\ee
where $\CM_{ii}$ is forward scattering amplitude,
$s$ is c.m. energy squared and $\sigma$ is total cross-section, $ep\to eX$.
The same holds true for contribution of each IS $h$
separately, i.e.
\be
 \Im \CM_{ii}^{(h)} = 2|p|\sqrt{s} \, \sigma^{(h)}
\ee
here $\sigma^{(h)}$ is the cross-section for $ep\to eh$.
Putting $u'=u$ and $U'=U$ in Eq.~(\ref{TPEampl}), we easily get
\be
 \Im \CM_{ii} = \frac{4\pi\alpha\nu}{Q^2} \Im \CG_E
\ee
thus the optical theorem implies
\be
 \Im \delta\CG_E^{(h)} > 0 \quad \text{for $Q^2\to 0$ at fixed $s$}
\ee
Note that for the elastic contribution such a check
constrains only infra-red divergent part.

We also have reproduced our results for target normal spin asymmetry
from Ref.~\cite{ourTNSA}, using Eq.~(\ref{TNSA}).
\begin{figure}
 \includegraphics[width=0.48\textwidth]{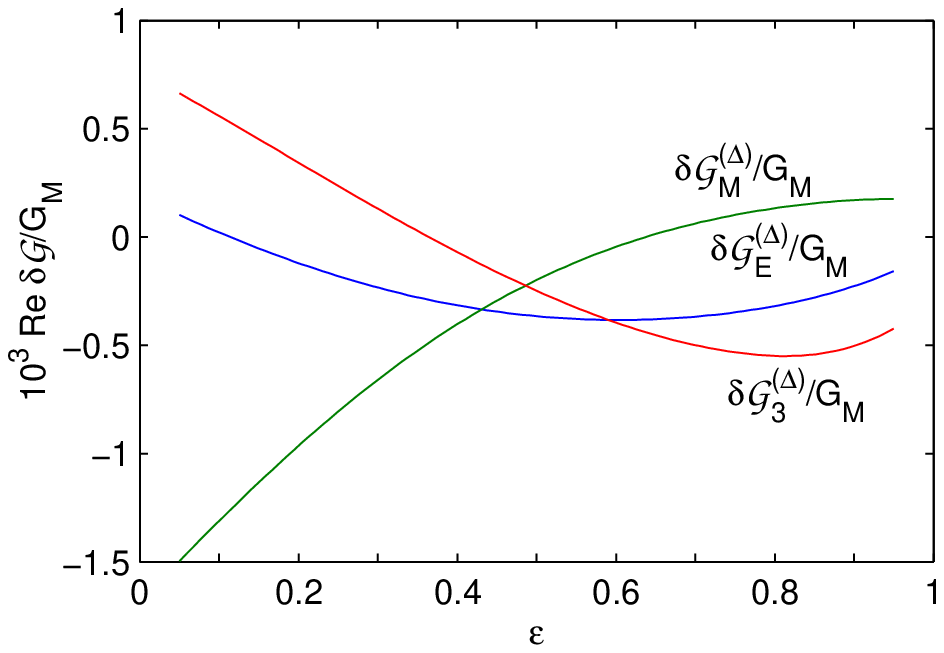}\quad
 \includegraphics[width=0.48\textwidth]{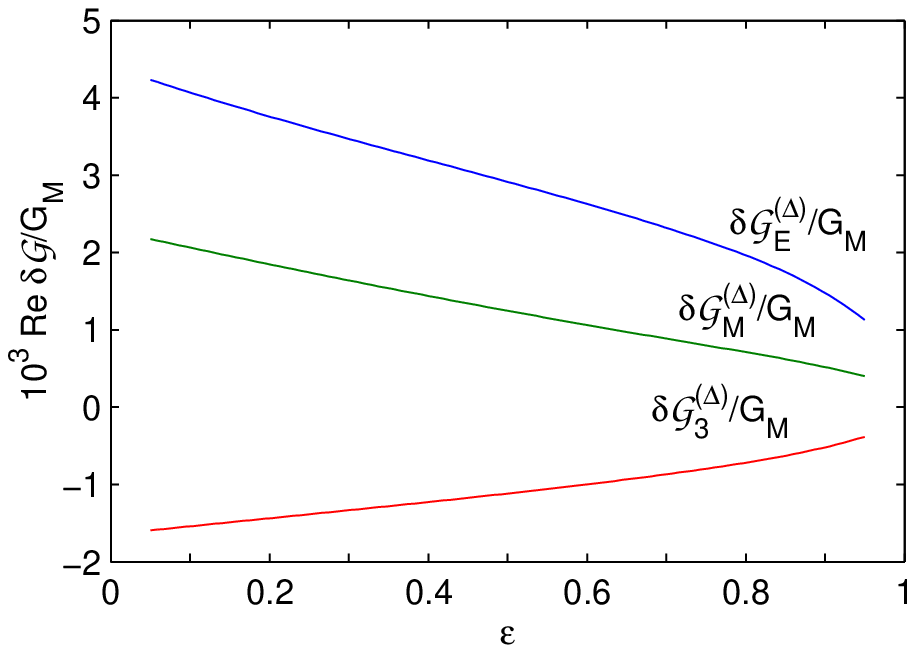}
 \caption{(Color online) TPE amplitudes at $Q^2=0.5\GeV^2$ (a) and $2\GeV^2$ (b)}
 \label{vs.epsilon}
\end{figure}

The $\ve$ dependence of the TPE amplitudes (real parts)
is shown in Fig.~\ref{vs.epsilon}.
It is substantially nonlinear for $Q^2<1\GeV^2$ (a)
and becomes almost linear for $Q^2>1\GeV^2$ (b) for all three amplitudes.
Recall that the elastic contribution $\delta\CG_M^{(el)}$,
is almost linear in $\ve$, and thanks to this fact
Rosenbluth plots remain linear even with the corresponding
correction taken into account.
It is interesting to study how much nonlinearity is introduced
by Delta contribution. We apply the method of Ref.~\cite{Tvaskis},
namely, we fit {\it calculated} OPE cross-section plus TPE correction
at fixed $Q^2$ by the quadratic function of $\ve$:
\be
 \sigma + \delta\sigma = P_0 \left[ 1 + P_1(\ve-0.5) + P_2 (\ve-0.5)^2 \right]
\ee
This gives us the nonlinearity coefficient $P_2$ as a function of $Q^2$.
The Delta contribution to nonlinearity coefficient turns out to be
rather small, $|P_2^{(\Delta)}|< 0.014 $ for $0.5\GeV^2<Q^2<5\GeV^2$,
whereas the total (elastic + Delta) contribution varies from 0.005 to $-0.06$.
This should be compared to experimental value $P_2 = 0.019\pm 0.027$ \cite{Tvaskis}.

In Fig.~\ref{dsig} we plot the TPE correction to the cross-section at 
$Q^2 = 1\GeV^2$ and $Q^2 = 3\GeV^2$.
This is the same quantity as in Fig.~2 of Ref.~\cite{BlundenDelta},
and is in qualitative agreement with the latter.
\begin{figure}
\parbox{0.49\textwidth}{
 \includegraphics[width=0.48\textwidth]{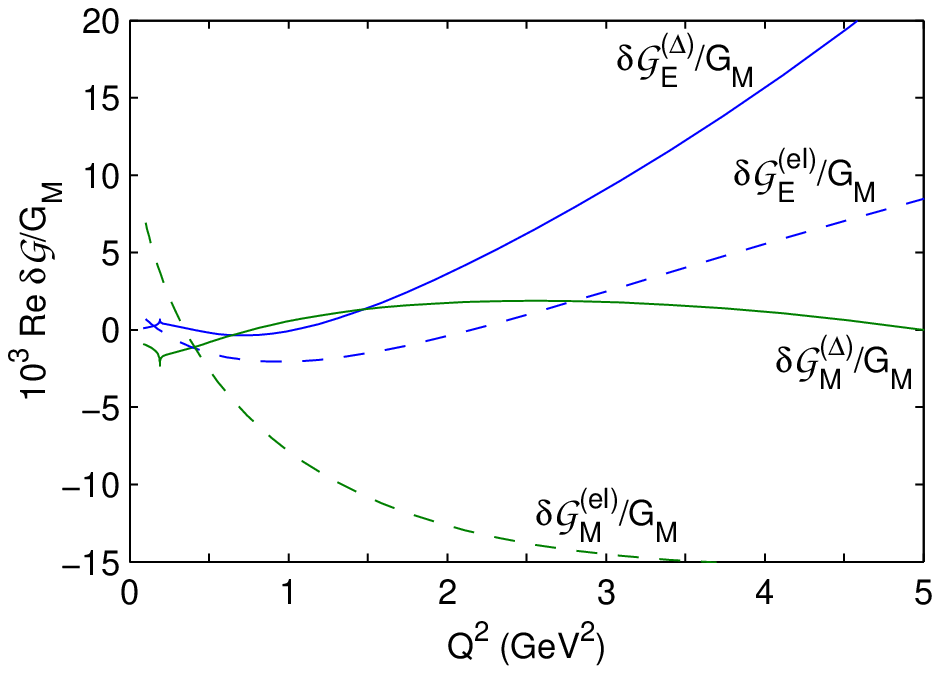}
 \caption{(Color online) Comparison of Delta (solid) and elastic (dashed) contributions,
   $\ve=0.25$.}\label{vs.elastic}
}\quad
\parbox{0.49\textwidth}{
 \includegraphics[width=0.48\textwidth]{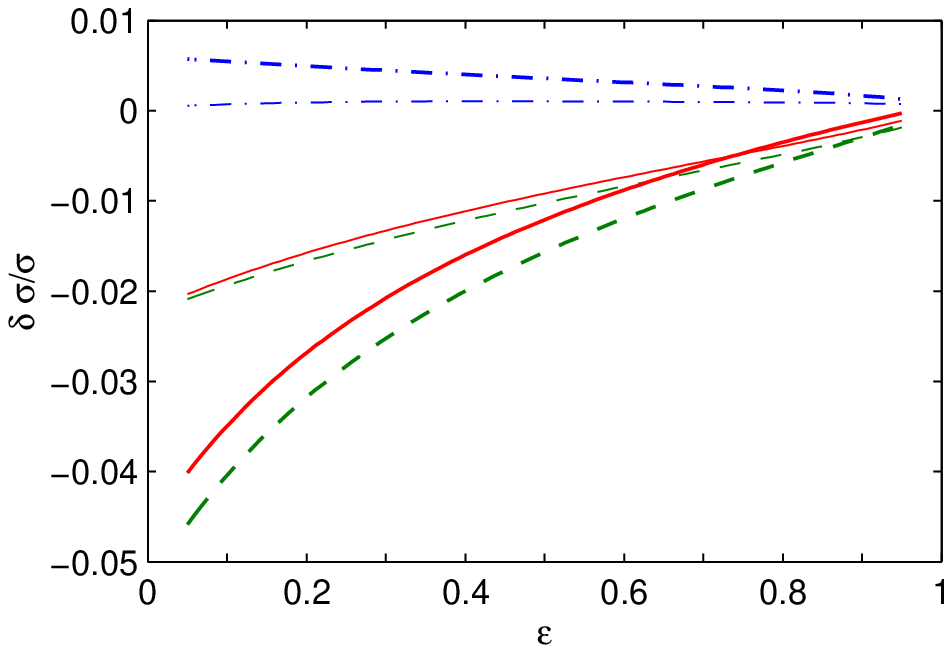}
 \caption{(Color online) TPE correction to cross-section at $Q^2=1\GeV^2$ (thin)
    and $3\GeV^2$ (thick).
    Proton contribution (dashed), Delta contribution (dash-dotted)
    and total (solid).}\label{dsig}
}
\end{figure}

\begin{figure}
\parbox{0.49\textwidth}{
 \includegraphics[width=0.48\textwidth]{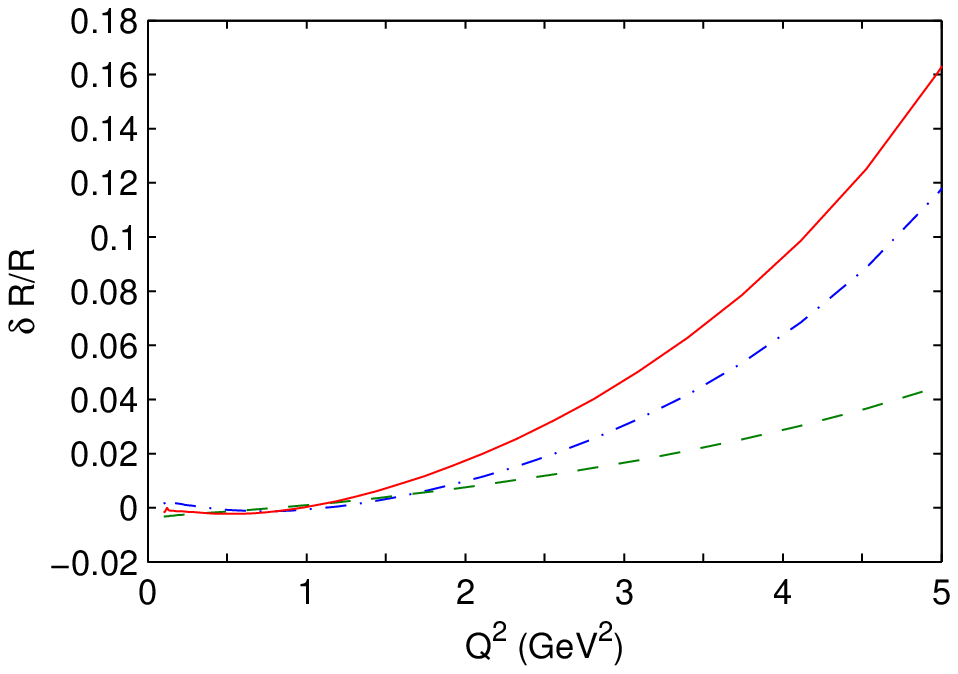}
 \caption{(Color online) TPE correction to measured form factor ratio at $\ve=0.5$.
    Proton contribution (dashed), Delta contribution (dash-dotted)
    and total (solid).}\label{deltaR}  
}\quad
\parbox{0.49\textwidth}{
 \includegraphics[width=0.48\textwidth]{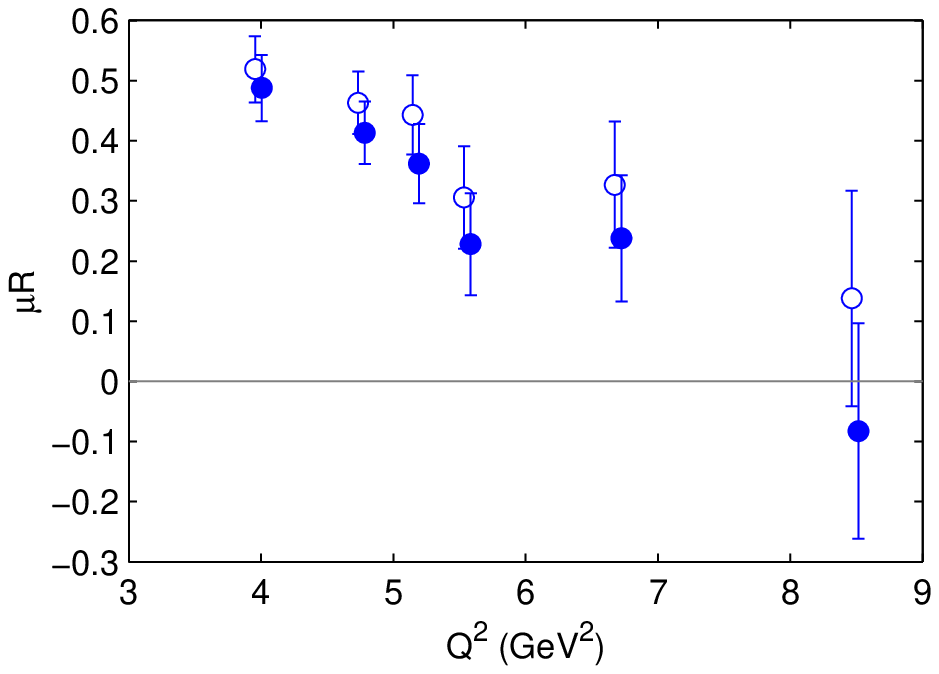}
 \caption{(Color online) Results of PT experiments, with (solid symbols)
    and without (hollow symbols) TPE correction.
    Points are slightly offset in $Q^2$ for clarity.}\label{PT+dR}
}
\end{figure}

As long as now we have individual TPE amplitudes,
we may easily obtain the TPE correction to FF ratio, Eq.~(\ref{dR}).
It is plotted in Fig.~\ref{deltaR}, for $\ve=0.5$.
Of course, one must keep in mind that this correction is also $\ve$ dependent,
as the TPE amplitudes are.
The correction grows rapidly as $Q^2\to \infty$ due to
\begin{itemize}
\item large amplitude $\delta\CG_E$, growing with $Q^2$,
\item smallness of FF ratio $R$ itself, as it tends to zero near $Q^2 \sim 10\GeV^2$.
\end{itemize}

In Table~\ref{Table1}, the total correction,
$\delta R = \delta R^{(el)} + \delta R^{(\Delta)}$,
is shown for the kinematical conditions of experiments \cite{Gayou,Puckett}.
As the correction is always much larger than the quoted systematic error,
it clearly needs to be taken into account
in polarization measurements at high $Q^2$.
With this TPE correction applied, the FF ratio becomes negative already
at $Q^2 = 8.5\GeV^2$ (Fig.~\ref{PT+dR}).
\begin{table}
\begin{tabular}{|ccc|c|c|}
\hline
Expt. & $Q^2$ & $\ve$ & $\mu R_{exp} \pm$stat.$\pm$syst. & $\mu \delta R$ \\
\hline
   & 3.98 & 0.71 & $0.517 \pm 0.055 \pm 0.009$ & 0.031 \\
\cite{Gayou}
   & 4.76 & 0.59 & $0.450 \pm 0.052 \pm 0.012$ & 0.050 \\
   & 5.56 & 0.45 & $0.354 \pm 0.085 \pm 0.019$ & 0.078 \\
\hline
   & 5.17 & 0.37 & $0.443 \pm 0.066 \pm 0.018$ & 0.081 \\
\cite{Puckett}
   & 6.70 & 0.51 & $0.327 \pm 0.105 \pm 0.022$ & 0.089 \\
   & 8.49 & 0.24 & $0.138 \pm 0.179 \pm 0.043$ & 0.221 \\
\hline
\end{tabular}
\caption{Form factor ratio, measured in PT experiments
  ($\mu R_{exp}$) and corresponding TPE corrections ($\mu\delta R$).
  The explicit factor of $\mu=2.793$ appears here
  as it is not included in our definition of $R$.}\label{Table1}
\end{table}
But is our scheme of calculation (elastic + Delta ISs)
perfectly adequate for $Q^2 \sim 5-10\GeV^2$?
We think that at least one needs to estimate the contributions
of other prominent resonances as well as multi-particle states
%
before we may apply the correction to data.
Whether these contributions are small? It is not quite clear.
Even if they are, there are many ISs that contribute,
and it is not clear what will be the total effect.

Of course, it would be nice to have a QCD-style calculation for this observable.
Unfortunately, this is a hard task.
Leading-twist QCD calculation yields only two amplitudes, $\CG_M$ and $\CG_3$,
since a virtual photon (gluon) cannot flip quark spin.
The calculation of electric FF $G_E$ or TPE amplitude $\delta\CG_E$
requires, at least, knowledge of quark transverse momenta distribution.

Summarizing, we believe that our results give a strong indication
that TPE corrections coming from the inelastic intermediate states
may be of great importance to polarization measurements at high $Q^2$,
and thus deserve further thorough investigation.

\end{document}